\documentclass[12pt]{article}
\usepackage[cp1251]{inputenc}
\usepackage{amsfonts}
\usepackage[english]{babel}
\usepackage{eufrak}
\usepackage{amssymb}

\begin{document}
 \sloppy
\linespread{1.6}
\title{Nonequlibrium Renormalization Theory III}
\author{D. V. Prokhorenko \footnote{Institute of Spectroscopy, RAS 142190 Moskow Region, Troitsk}}\maketitle
\begin{abstract}
In the present paper we develop the general theory of
renormalization of some nonequilibrium diagram technique. This
technique roughly is the Keldysh diagram technique. To develop our
theory we have used the Bogoliubov --- Parasiuk \(R\) - operation
method. Our theory can be used for the studing of the divergences
in kinetic equations.
\end{abstract}
\newpage
\section{Introduction}
The main goal of this paper is to prove that for the general class
of two-particle interaction there exist (in the sense of formal
power series on coupling constant) the stationary translation
invariant states on the algebra of canonical commutative relations
which satisfy the weak cluster property and can not be described
by the Gibbs formula. To solve this problem we use the
\(R\)-operation method \cite{1,2,3}. The methods of
\(R\)-operation for studying the large time behavior of quantum
systems have been used in \cite{4}. The philosophical consequences
of our statement have been discussed in our first paper on this
topic. This statement follows from the theorem --- construction of
the last section of this paper.
\section{The Algebra of Canonical Commutative Relations}
Let \(S( \mathbf{R}^3)\) be a Schwatrz space of test functions
(infinitely-differentiable functions decaying at infinity faster
than any inverse polynomial with all its derivatives). The algebra
of canonical commutative relations \(\mathcal{C}\) is an unital
algebra generated by symbols \(a^+(f)\) and \(a(f)\) \(f \in S(
\mathbf{R}^3)\) satisfying the following canonical commutative
relations.

a) \(a^+(f)\) is a linear functional of \(f\),

b) \(a(f)\) is an antilinear functional of \(f\),

\begin{eqnarray}
{[a(f),a(g)]}={[a^+(f),a^+(f)]}=0,\nonumber\\
{[a(f),a^+(g)]}=\langle f,g\rangle,
\end{eqnarray}
where \( \langle f,g\rangle\) is a standard scalar product in
\(L^2(\mathbf{R}^3)\),
\begin{eqnarray}
\langle f,g\rangle:=\int d^3 x f^*(x)g(x).
\end{eqnarray}
Let \(\rho_0\) be a Gauss state on \(\mathcal{C}\) defined by the
following correlator
\begin{eqnarray}
\rho_0(a^+(k)a^+(k'))=\rho_0(a(k)a(k'))=0,\nonumber\\
\rho_0(a^+(k)a(k'))=n(k)\delta(k-k'),
\end{eqnarray}
where \(n(k)\) is a real-valued function from the Schwartz space.
In the case then
\begin{eqnarray}
n(k)=\frac{e^{-\beta(\omega(k)-\mu)}}{1-e^{-\beta(\omega(k)-\mu)}},
\end{eqnarray}
where \(\mu \in \mathbf{R}\), \(\mu<0\), \(\rho_0\) is called the
Plank state. Here \(\omega(k)=\frac{k^2}{2}\).

Let \(\mathcal{C}'\) be a space of linear functionals on
\(\mathcal{C}\), and \(\mathcal{C}'_{+,1}\) be a set of all states
on \(\mathcal{C}\). Let us make the GNS construction corresponding
to the algebra \(\mathcal{C}\) and the Gauss state \(\rho_0\). We
obtain the set \((\mathcal{H},D,\hat{},\rangle)\) consisting of
the Hilbert space \(\mathcal{H}\), the dense linear subspace \(D\)
in \(\mathcal{H}\), the representation \(\hat{}\) of
\(\mathcal{C}\) by linear operators from \(D\) to \(D\), and
cyclic vector \(\rangle \in D \), i.e. the vector such that
\(\hat{C}\rangle=D\). This set satisfy to the following condition:
\(\forall a \in \mathcal{C}\) \(\langle\hat{a}\rangle=\rho_0(a)\).
Below we will omit the symbol \(\hat{}\), i.e. we will write \(a\)
instead of \(\hat{a}\).

Let us introduce the field operators:
\begin{eqnarray}
\Psi(x)=\frac{1}{(2\pi)^{\frac{3}{2}}} \int
e^{ikx}a(k)dk,\nonumber\\
\Psi^+(x)=\frac{1}{(2\pi)^{\frac{3}{2}}} \int e^{-ikx}a^+(k)dk.
\end{eqnarray}

We say that the state \(\rho\) on \(\mathcal{C}\) satisfy to the
weak cluster property if

\begin{eqnarray}
\lim_{{a}\rightarrow\infty}\int
\langle\Psi^\pm(t,{x}_1+\delta_1e_1{a})
...\Psi^\pm(t,{x}_n+\delta_n e_1{a})\rangle f({x}_1,...,{x}_n)d^3x_1...d^3x_n\nonumber\\
=\int\langle\Psi^\pm(t,{x}_{i_1})...\Psi^\pm(t,{x}_{i_k})\rangle
\langle\Psi^\pm(t,{x}_{i_k})...\Psi^\pm(t,{x}_{i_n})\rangle
\nonumber\\
\times f({x}_1,...,{x}_n)d^3x_1...d^3x_n,
\end{eqnarray}
where \(\delta_i\in\{1,0\},\;i=1,2...n\) and
\begin{eqnarray}
i_1<i_2<...<i_k,\nonumber\\
i_{k+1}<i_{k+2}<...<i_n,\nonumber\\
\{i_1,i_2,...,i_k\}=\{i=1,2...n|\delta_i=0\}\neq\emptyset,\nonumber\\
\{i_{k+1},i_{k+2},...,i_n\}=\{i=1,2...n|\delta_i=1\}\neq\emptyset.
\end{eqnarray}
\(f({x}_1,...,{x}_n)\) is a test function. \(e_1\) is an unite vector parallel to the \(x\)-axis.

 \textbf{Definition} The vector of
the form
\begin{eqnarray}
\int v(p_1,...,p_n)a^{\pm}(p_1)...a^{\pm}(p_n)\rangle
d^3p_1...d^3p_n,\nonumber\\
v(p_1,...,p_n) \in S(\mathbb{R}^{3n}). \label{1}
\end{eqnarray}
is called a finite vector. The finite linear combination of the
vectors of the form (\ref{1}) is also called a finite vector.

Let \(f(x_1,...,x_k|y_1,...,y_l|v_1,...,v_m|w_1,...,w_n)\) be a
function of the form
\begin{eqnarray}
f(x_1,...,x_k|y_1,...,y_l|v_1,...,v_m|w_1,...,w_n)\nonumber\\
=g(x_1,...,x_k|y_1,...,y_l|v_1,...,v_m|w_1,...,w_n)\nonumber\\
\times\delta(\sum \limits_{i=1}^k x_i+\sum\limits_{j=1}^m w_j-
\sum\limits_{f=1}^l y_f-\sum\limits_{g=1}^n v_g),
\end{eqnarray}
where \(g\) is a function from Schwartz space.

Consider the following functional on \(\mathcal{C}\)
\begin{eqnarray}
\rho_f(A):=\int \prod \limits_{i=1}^k dx_i\prod \limits_{j=1}^l
dx_j \prod \limits_{f=1}^m dv_f \prod \limits_{g=1}^n dw_g \nonumber\\
\times f(x_1,...,x_n|y_1,...,y_l|v_1,...,v_m|w_1,...,w_n)\nonumber\\
\times
\rho_0(:a(x_1)...a(x_n)\overbrace{a^+(y_1)...a^+(y_l):A:a(v_1)...a(v_m)}a^+(w_1)...a^+(w_n):).
\label{7}
\end{eqnarray}
Here the symbol
\begin{eqnarray}
:(.\overbrace{.):A:(.}.):
\end{eqnarray}
means that when one transform the previous expression to the
normal form according to the Gauss property of \(\rho_0\) one must
neglects by all correlators \(\rho(a^\pm(x_1)a^\pm(x_n)\) such
that \(a^\pm(x_1)\) and \(a^\pm(x_n)\) booth do not
came from \(A\).

Let \(\widetilde{\mathcal{C}'}\) be a subspace in \(\mathcal{C'}\)
spanned on the states just defined.

Now let us introduce an useful method for the representation of
the states just defined.

Let \(\mathcal{C}_2=\mathcal{C}_+\otimes\mathcal{C}_-\), where
\(\mathcal{C}_+\) and \(\mathcal{C}_+\) are the algebras of
canonical commutative relations. The algebras
\(\mathcal{C}_{\pm}\) are generated by the generators
\(a_\pm(k),a^+_\pm(k)\) respectively satisfying to the following
relations:
\begin{eqnarray}
{[a^+_+(k),a^+_+(k')]}={[a_+(k),a_+(k')]}=0,\nonumber\\
{[a^+_-(k),a^+_-(k')]}={[a_-(k),a_-(k')]}=0,\nonumber\\
{[a_+(k),a^+_+(k')]}=\delta(k-k'),\nonumber\\
{[a_-(k),a^+_-(k')]}=\delta(k-k'),\nonumber\\
{[a^{\pm}_+(k),a^{\pm}_-(k)]}=0.
\end{eqnarray}
Here we put by definition \(a^-_{\pm}:=a_{\pm}\).
Let us consider the following Gauss state \(\rho_0'\) on
\(\mathcal{C}_2\) defined by its two-point correlator
\begin{eqnarray}
\rho_0'(a^{\pm}_-(k)a^{\pm}_-(k'))=\rho_0(a^{\pm}(k)a^{\pm}(k')),\nonumber\\
\rho_0'(a^{\pm}_+(k)a^{\pm}_+(k'))=\rho_0(a^{\mp}(k')a^{\mp}(k)),\nonumber\\
\rho_0'(a^+_+(k)a^-_-(k'))=\rho_0'(a^-_+(k)a^+_-(k'))=0, \nonumber\\
\rho_0'(a^-_+(k)a^-_-(k'))=n(k)\delta(k-k'),\nonumber\\
\rho_0'(a^+_+(k)a^+_-(k'))=(1+n(k))\delta(k-k').
\end{eqnarray}
Let us make the GNS construction corresponding to the state
\(\rho_0'\) and the algebra \(\mathcal{C}_2\). We obtain the set
\((\mathcal{H}',\tilde{D},\hat{},\rangle)\) consisting of the
Hilbert space \(\mathcal{H}'\), the dense linear subspace
\(\tilde{D}\) in \(\mathcal{H}'\), the representation \(\hat{}\)
of \(\mathcal{C}_2\) by means linear operators from \(\tilde{D}\)
to \(\tilde{D}\), and the cyclic vector \(\rangle \in \tilde{D}
\), i.e. the vector such that \(\hat{C}\rangle=\tilde{D}\). This
set satisfy the following condition: \(\forall a \in \mathcal{C}\)
\(\langle\hat{a}\rangle=\rho_0(a)\). Below we will omit the symbol
\(\hat{}\), i.e. we will write \(a\) instead of \(\hat{a}\).

Now we can rewrite the functional, defined in (\ref{7})
\(\rho_f\) as follows
\begin{eqnarray}
\rho_f(A)=\langle A'S_f\rangle,
\end{eqnarray}
where \(A'\) is an element of \(\mathcal{C}_2\) such that it
contains only the operators \(a_-,a_-^+\) and can be represented
from \(a_-,a_-^+\) as the same way as \(A\) can be represented
from \(a,a^+\). \(S_f\) is an element of \(\mathcal{C}_2\) of the
form
\begin{eqnarray}
S_f=\int \prod \limits_{i=1}^k dx_i\prod \limits_{j=1}^l
dx_j \prod \limits_{f=1}^m dv_f \prod \limits_{g=1}^n dw_g \nonumber\\
\times f(x_1,...,x_n|y_1,...,y_l|v_1,...,v_m|w_1,...,w_n)\nonumber\\
\times
:a^+_+(x_1)...a^+_+(x_n)a_+(y_1)...a_+(y_l)a_-(v_1)...a_-(v_m)a^+_-(w_1)...a^+_-(w_n):.
\label{2}
\end{eqnarray}
Here the symbol \(:...:\) is a normal ordering with respect to the
state \(\rho_0'\).

Denote by \(\tilde{D}'\) the space dual to  \(\tilde{D}\). We just
construct the injection from \(C'\) into \(\tilde{D}'\). Denote
its image by \(\tilde{\mathcal{H}'}\).

By definition the space \(\mathcal{C}''\) is a space of all
functionals on \(\mathcal{C}\) which can be represented as finite
linear combinations of the following states
\begin{eqnarray}
\rho(A)=\langle A':S_{f_1}...S_{f_n}:\rangle.
\end{eqnarray}
 Here \(A'\) is an element of \(\mathcal{C}_2\) such that it contains only the
 operators \(a_-,a_-^+\) and can be represented from \(a_-,a_-^+\) as the
same way as \(A\) can be represented from \(a,a^+\) and
\(S_{f_i}\) are the elements of the form (\ref{2}). Denote by
\(\tilde{\mathcal{H}}''\) the subspace in \(\tilde{D}'\) spanned
on the vectors \(:S_{f_1}...S_{f_n}:\rangle\).

There exists an involution \(\star\) on \(\tilde{H}'\) defined by
the following formula:
\begin{eqnarray}
\{\int \prod \limits_{i=1}^k dx_i\prod \limits_{j=1}^l
dx_j \prod \limits_{f=1}^m dv_f \prod \limits_{g=1}^n dw_g \nonumber\\
\times f(x_1,...,x_n|y_1,...,y_l|v_1,...,v_m|w_1,...,w_n)\nonumber\\
\times
:a^+_+(x_1)...a^+_+(x_n)a_+(y_1)...a_+(y_l)a_-(v_1)...a_-(v_m)a^+_-(w_1)...a^+_-(w_n):\rangle\}^*
\nonumber\\
 =\int \prod \limits_{i=1}^k dx_i\prod \limits_{j=1}^l
dx_j \prod \limits_{f=1}^m dv_f \prod \limits_{g=1}^n dw_g \nonumber\\
\times f^*(x_1,...,x_n|y_1,...,y_l|v_1,...,v_m|w_1,...,w_n)\nonumber\\
\times
:a^+_-(x_1)...a^+_-(x_n)a_-(y_1)...a_-(y_l)a_+(v_1)...a_+(v_m)a^+_+(w_1)...a^+_+(w_n):\rangle.
\end{eqnarray}

We define the involution \(\star\) on \({\rm Hom \mit}
(\tilde{H}',\tilde{H}')\) by the following equation:
\begin{eqnarray}
(a |f\rangle)^\star=a^\star(|f\rangle)^*,
\end{eqnarray}
where \(a \in {\rm Hom \mit} (\tilde{H}',\tilde{H}')\) and
\(|f\rangle \in \tilde{H}'\).

We define also the involution \(\star\) on \(\mathcal{C}^2\) by
the following equation:
\begin{eqnarray}
\{\int \prod \limits_{i=1}^k dx_i\prod \limits_{j=1}^l
dx_j \prod \limits_{f=1}^m dv_f \prod \limits_{g=1}^n dw_g \nonumber\\
\times f(x_1,...,x_n|y_1,...,y_l|v_1,...,v_m|w_1,...,w_n)\nonumber\\
\times
:a^+_+(x_1)...a^+_+(x_n)a_+(y_1)...a_+(y_l)a_-(v_1)...a_-(v_m)a^+_-(w_1)...a^+_-(w_n):\}^*
\nonumber\\ =\int \prod \limits_{i=1}^k dx_i\prod \limits_{j=1}^l
dx_j \prod \limits_{f=1}^m dv_f \prod \limits_{g=1}^n dw_g \nonumber\\
\times f^*(x_1,...,x_n|y_1,...,y_l|v_1,...,v_m|w_1,...,w_n)\nonumber\\
\times
:a^+_-(x_1)...a^+_-(x_n)a_-(y_1)...a_-(y_l)a_+(v_1)...a_+(v_m)a^+_+(w_1)...a^+_+(w_n):,
\end{eqnarray}
where \(f(x_1,...,x_k|y_1,...,y_l|v_1,...,v_m|w_1,...,w_n)\) be a
test function of its arguments. Note that the involution on \({\rm
Hom \mit} (\tilde{H}',\tilde{H}')\) extends the involution on
\(\mathcal{C}^2\). We say that the element \(a \in \mathcal{C}_2\)
is real if \(a^\star=a\). The involution on \(\tilde{H}''\) can be
defined by the similar way.

\section{The von Neumann Dynamics}
Suppose that our system described by the following Hamiltonian
\begin{eqnarray}
H=H_0+\lambda V,
\end{eqnarray}
where
\begin{eqnarray}
H_0=\int d^3k (\omega(k)-\mu)a^+(k)a(k)\; \rm and \mit \nonumber\\
V=\int d^3p_1d^3p_2d^3q_1d^3q_2 v(p_1,p_2|q_1,q_2)\nonumber\\
\times\delta(p_1+p_2-q_1-q_2)a^+(p_1)a^+(p_2)a(q_1)a(q_2).
\end{eqnarray}
Here the kernel \(v(p_1,p_2|q_1,q_2)\) belongs to the Schwartz
space of test functions. To point out the fact that \(H\) is
represented through the operators \(a^+,\;a^-\) we will write \(
H(a^+,a^-)\).

The von Neumann dynamics takes place in the space
\(\tilde{\mathcal{H}}''\) and defined by the following
differential equation:
\begin{eqnarray}
\frac{d}{dt}|f\rangle=\mathcal{L}|f\rangle,
\end{eqnarray}
where the von Neumann operator has the form
\begin{eqnarray}
\mathcal{L}=-iH(a^+_-,a^-_-)+iH^\dagger(a^+_+,a^-_+),
\end{eqnarray}
where we put by definition:
\begin{eqnarray}
(\int \prod \limits_{i=1}^n d p_i \prod \limits_{j=1}^m d q_j
v(p_1,...,p_n|q_1,...,q_m):a^+(p_1)...a^+(p_n)a(q_1)...a(q_n):)^\dagger\nonumber\\
=\int \prod \limits_{i=1}^n d p_i \prod \limits_{j=1}^m d q_j
v(p_1,...,p_n|q_1,...,q_m)^\ast:a^+(p_1)...a^+(p_n)a(q_1)...a(q_n):.
\end{eqnarray}
Let us divide the von Neumann operator into the free operator
\(\mathcal{L}\) and the interaction \(\mathcal{L}_{int}\),
\(\mathcal{L}=\mathcal{L}_0+\lambda \mathcal{L}_{int}\), where
\begin{eqnarray}
\mathcal{L}_0=-iH_0(a^+_-,a^-_-)+iH_0^\dagger(a^+_+,a^-_+),\nonumber\\
\mathcal{L}_{int}=-iH_{int}(a^+_-,a^-_-)+iH_{int}^\dagger(a^+_+,a^-_+).
\end{eqnarray}
Note that the operators \(\mathcal{L}_0\) and \(\mathcal{L}_1\)
are real.

 Let us introduce kinetic evolution operator (in the
interaction representation)
\begin{eqnarray}
U(t'',t')=e^{-\mathcal{L}_0t''}e^{\mathcal{L}(t''-t')}e^{\mathcal{L}_0t'}.
\end{eqnarray}
After differentiating on \(t\) we find the differential equation
for \(U(t,t')\).
\begin{eqnarray}
\frac{d}{dt}U(t,t')=\mathcal{L}_{int}(t)U(t,t'),
\end{eqnarray}
where
\begin{eqnarray}
\mathcal{L}_{int}(t)=e^{-\mathcal{L}_0t}\mathcal{L}_{int}e^{\mathcal{L}_0t}.
\end{eqnarray}
So the state \(\rangle_{\rho}\) under consideration in the
interaction representation in the space \(\tilde{\mathcal{H}}'' \)
has the form
\begin{eqnarray}
\rangle_\rho=T \rm exp \mit (\int \limits_{-\infty}^0
\mathcal{L}_{int}(t)dt)\rangle,
\end{eqnarray}
where \(T\) is the time-ordering operator.

\section{Dynamics of Correlations}
Let us construct some new representation of dynamics useful for
the renormalization program. This representation is called the
dynamics of correlations. The ideas of the dynamics of correlations belongs to I.R. Prigogin \cite{Pr}.
 The dynamics of correlations take place
in the space
\begin{eqnarray}
\mathcal{H}_c:={\bigoplus \limits_0^{\infty} \rm sym \mit
{\otimes}^n} \tilde{\mathcal{H}}'.
\end{eqnarray}
Now let us describe how the operators \(\mathcal{L}_0^c\) and
\(\mathcal{L}_{int}^c\) acts in the space \(\mathcal{H}_c\).

Let us define the action of operators \(\mathcal{L}_0^c\) and
\(\mathcal{L}_{int}^c\) which are corresponds to the operators
\(\mathcal{L}_0\) and \(\mathcal{L}_{int}\).

By definition all the space \(\otimes^n \tilde{{\mathcal{H}}}'\)
are invariant under the action of operators \(\mathcal{L}_{0}^c\).
Note that the space \(\tilde{{\mathcal{H}}}'\) is invariant under
the action of operator \(\mathcal{L}_{0}\). Let us denote the
restriction of \(\mathcal{L}_{0}\) to the space
\(\tilde{{\mathcal{H}}}'\) by the symbol \(\mathcal{L}_{0}'\). By
definition the restriction  of \(\mathcal{L}_{0}^c\) to the each
subspace \(\rm sym \mit\otimes^n \tilde{{\mathcal{H}}}'\) of
\(\mathcal{H}_c\) has the form
\begin{eqnarray}
\mathcal{L}'_0\otimes\mathbf{1}\otimes...\otimes\mathbf{1}+
\mathbf{1}\otimes\mathcal{L}'_0\otimes...\otimes\mathbf{1}+...+
\mathbf{1}\otimes\mathbf{1}\otimes...\otimes\mathcal{L}'_0.
\end{eqnarray}
Now let us define \(\mathcal{L}^c_{int}\). Let \(|f\rangle \in
\mathcal{H}_c\), belongs to the subspace \( \otimes^n
\tilde{{\mathcal{H}}}'\) and has the form:
\begin{eqnarray}
|f\rangle=\sum \limits_{i=0}^m f_1^i\rangle\otimes...\otimes
f_n^i\rangle,
\end{eqnarray}
where \(f_j^i\) has the form
\begin{eqnarray}
f_i^j=\int \prod \limits_{i=1}^k dx_i\prod \limits_{j=1}^l
dx_j \prod \limits_{f=1}^m dv_f \prod \limits_{g=1}^n dw_g \nonumber\\
\times f(x_1,...,x_n|y_1,...,y_l|v_1,...,v_m|w_1,...,w_n)\nonumber\\
\times
:a^+_+(x_1)...a^+_+(x_n)a_+(y_1)...a_+(y_l)a_-(v_1)...a_-(v_m)a^+_-(w_1)...a^+_-(w_n):.\label{35}
\end{eqnarray}

By definition,
\begin{eqnarray}
\mathcal{L}_{int}^{c,l}|f\rangle=0
\end{eqnarray}
if \(l> n\). Let us consider the following vector in
\(\tilde{\mathcal{H}}''\)
\begin{eqnarray}
\sum \limits_{i=1}^{m}:\prod \limits_{j=1}^n f^i_j:\rangle.
\end{eqnarray}
Let us transform the expression \(\mathcal{L}_{int} \sum
\limits_{i=1}^m :\prod \limits_{j=1}^n f^i_j:\) to the normal
form. Let us denote by \(h_l\) the sum of all the terms in the
previous expression such that precisely \(l\) operators \(f^i_j\)
couples with \(\mathcal{L}_{int}\).
 We find that \(h_l\rangle\) has the following form
\begin{eqnarray}
h_l\rangle=\sum \limits_{i=1}^k :g_1^i...g_{n-l+1}^i:\rangle
\end{eqnarray}
for some \(k\). Here \(g^i_k\) has the form of right hand side of (\ref{35}).
Now let us consider the following vector
\begin{eqnarray}
|f\rangle_l^c=\rm sym \mit \sum \limits_{i=1}^k
:g_1^i:\rangle\otimes...\otimes :g^i_{n-l+1}:\rangle,
\end{eqnarray}
where we define symmetrization operator as follows
\begin{eqnarray}
\rm sym \mit (f_1\otimes...\otimes f_n)\nonumber\\
=\frac{1}{n!}\sum \limits_{\sigma \in S_n}
f_{\sigma_1}\otimes...\otimes f_{\sigma(n)}.
\end{eqnarray}
(\(S_n\) --- the group of permutation of \(n\) elements.) Put by
definition
\begin{eqnarray}
\mathcal{L}_{int}^{c,l}|f\rangle=|f\rangle_l^c.
\end{eqnarray}
Analogously, in the following expression
\begin{eqnarray}
\mathcal{L}_{int}\sum \limits_{i=1}^m :\prod \limits_{j=1}^n
f^i_j:\rangle
\end{eqnarray}
let us keep only the terms such that \(\mathcal{L}_{int}\) do not
couples with any of \(f^i_j\). Let us write the sum of such terms
as follows
\begin{eqnarray}
\sum \limits_{i=1}^f :\prod \limits_{j=1}^{n+1} h^i_j:\rangle.
\end{eqnarray}
Here \(h^i_j\rangle\) has the form of right hand side of (\ref{35}).
Let \(|h\rangle\) be a vector in \( {\rm sym \mit \otimes^{n+1}}
\tilde{\mathcal{H}}'\) defined as follows
\begin{eqnarray}
|h\rangle=\rm sym \mit \sum \limits_{i=1}^f
\bigotimes\limits_{j=1}^{n+1}:h^i_j:\rangle.
\end{eqnarray}
Put by definition
\begin{eqnarray}
\mathcal{L}_{int}^{c,0}|f\rangle=  |h\rangle.
\end{eqnarray}

We have the evident linear map
\(F:\mathcal{H}_c\rightarrow\tilde{\mathcal{H}}''\) which assigns
to each vector \(\rm sym \mit
:f_1:\rangle\otimes...\otimes:f_n:\rangle\) the vector
\(:f_1...f_n:\rangle\). Denote by \(U^c\) the evolution operator
in interaction representation in the dynamics of correlation. The
following relation holds:
\begin{eqnarray}
F\circ U^c(t',t'')=U(t',t'')\circ F.
\end{eqnarray}
\section{The correlations tree}
The useful representation of dynamics in \(\mathcal{H}_c\) is a
decomposition by so called the correlation's trees.

\textbf{Definition.} A graph is a triple \(T=(V,R,f)\), where
\(V\), \(R\) are finite sets called the set of vertices and lines
respectively and \(f\) is a map:
\begin{eqnarray}
h:R\rightarrow V^{(2)}\cup V\times \{+\}\cup V\times\{-\},
\end{eqnarray}
where \(V^{(2)}\) is a set of all disordered pairs \((v_1,v_2)\),
\(v_1, v_2 \in V\), \(v_1\neq v_2\).

If \((v_1,v_2)=f(r)\) for some \(r \in R\) we say that the
vertices \(v_1\) and \(v_2\) are connected by a line \(r\). If
\(f(r)=(v_1,v_2)\), \(v_1,v_2 \in V\) we say that the line \(r\)
is internal.

\textbf{Remark.} We use this non-usual definition of graphs only in purpose
of this section to simplify our notations.

\textbf{Definition.} The graph \(\Gamma\) is called connected if
for two any vertices  \(v,v'\) there exists a sequence of vertices
\(v=v_0,v_1,...,v_n=v'\) such that
 \(\forall\, i=0,...,n-1\) the vertices
 \(v_i\) and \(v_{i+1}\) are connected by some line.

By definition we say that the line \(r\) is an internal line if
\(f(r)=(v_1,v_2)\) for some vertices \(v_1\) and \(v_2\).

For each graph \(\Gamma\) we define its connected components by
the obvious way.

\textbf{Definition.}  We say that the graph \(\Gamma\) is a tree
or an acyclic graph if the number of its connected components
increases after removing an arbitrary line.

\textbf{Definition.} The elements of the set
\(f^{-1}(V\times\{-\})\) we call the shoots. Put by definition
\(R_{sh}=f^{-1}(V\times\{-\})\). The elements of the set
\(f^{-1}(V\times\{+\})\) we call the roots. Put by definition
\(R_{root}=f^{-1}(V\times\{+\})\).

\textbf{Definition.} Directed tree is triple \((T,
\Phi_v,\Phi_{sh})\), where \(T\) is a tree and \(\Phi_v\) and
\(\Phi_{sh}\) are the following maps:
\begin{eqnarray}
\Phi_v: V\rightarrow \{1,2,...\sharp V\},\nonumber\\
 \Phi_{sh}: R_{sh}\rightarrow\{1,2,...,\sharp R_{sh}\}.
\end{eqnarray}

\textbf{Definition.} We will consider the following two directed
trees \((T,\Phi_v,\Phi_{sh})\) and \((T',\Phi'_v,\Phi'_{sh})\) as
identical if we can identify the sets of lines \(R\) and \(R'\) of
\(T\) and \(T'\) respectively and identify the sets of vertices
\(V\) and \(V'\) of \(T\) and \(T'\) respectively such that after
these identification the trees \(T\) and \(T'\) become the same,
the functions \(\Phi_v\) and \(\Phi'_v\) become the same and the
functions \(\Phi_{sh}\) and \(\Phi'_{sh}\) become the same.

Denote by \(r(T)\) the number of roots of \(T\) and by \(s(T)\)
the number of shoots of \(T\). Below, we will denote each directed
tree \((T,\Phi_v,\Phi_{sh})\) by the same symbol \(T\) as a tree
omitting the referring to \(\Phi_v\), \(\Phi_{sh}\) and write
simply tree instead of the directed tree.

We say that the connected directed tree \(T\) is right if there
exists precisely one line from \(f^{-1}(V\times\{+\})\).

We say that the tree \(T\) is right if each its connected
component is right.

The vertex \(v\) of the tree \(T\) is called a root vertex if
\((v,+) \in f^{-1}(R)\).

To point out the fact that some object \(A\) corresponds to a tree
\(T\) we will often write \(A_T\). For example we will write
\(T=(V_T,R_T,f_T)\) instead of \(T=(V,R,f)\).

\textbf{Definition.} For each right tree \(T\) there exist an
essential partial ordering on the set of its vertices. Let us
describe it by induction on the number of its vertices. Suppose
that we have defined this relation for all right trees such that
the number of their vertices less or equal than \(n-1\). Let \(T\)
be a right tree such that the number of its vertices is equal to
\(n\). Let \(v_{max}\) be a root vertex of \(T\). Put by
definition that the vertex \(v_{max}\) is a maximal vertex. Let
\(v_1,...,v_k\) be all its children i.e. the vertices connected
with \(v_{max}\) by lines. By definition each vertex
\(v_i<v_{max},i=1,...,k\). We can consider the vertices
\(v_1,...,v_k\) as a root vertices of some directed trees
\(T_i,i=1,...,k\). By definition the set of vertices of \(T_i\)
consists of all vertices \(v\) which can be connected with \(v_i\)
by some path \(v=v'_1,....,v'_l=v_i\) such that \(v_{max}\neq
v'_j\) for all \(j=1,...,l\). The incident relations on \(T_i\)
are induced by incident relations on \(T\). Put by definition that
\(\forall (i,j),\;i,j=1,...,k,\;i\neq j\) and for all two vertices
\(v'_1 \in T_i\) and \(v'_2 \in T_j\) \(v'_1\nless v'_2\). If
\(v_1',v_2' \in T_i\) for some \(T_i\) we put \(v_1'\lessgtr v_2'
\) in \(T\) if and only if \(v_1'\lessgtr v_2'\) in the sense of
ordering on \(T_i\). We put also \(v<v_{max}\) for all vertex
\(v\neq v_{max}\). These relations are enough to define the
partial ordering on \(T\).

Below without loss of generality we suppose that for each the tree
of correlation \(T\) and its line \(r\) the pair
\((v_1,v_2)=f(r)\) satisfy to the inequality \(v_1>v_2\).

\textbf{Definition.} The tree of correlations \(C\) is a triple
\(C=(T,\varphi,\vec{\tau})\), where \(T\) is a directed tree,
\(\vec{\tau}\) is a map from \(R\setminus R_{sh}\) to
\(\mathbb{R}^+:=\{x \in \mathbb{R}|x\geq 0\}\):
\begin{eqnarray}
\vec{\tau}:R\setminus R_{sh}\rightarrow \mathbb{R}^+,\nonumber\\
r\mapsto \tau(r),\nonumber\\
(\tau(r))_{r \in \mathbb{R}}=\vec{\tau}(r),
\end{eqnarray}
and \(\varphi\) is a map which assigns to each vertex \(v\) of
\(T\) an element
\begin{eqnarray}
\varphi(v) \in {\rm Hom \mit} ({\bigotimes \limits_{r\rightarrow
v}} {\tilde{\mathcal{H}}', \tilde{\mathcal{H}}'})
\end{eqnarray}
of a space of linear maps from \(\bigotimes \limits_{r \rightarrow
v} \tilde{\mathcal{H}}'\) to \(\tilde{\mathcal{H}}'\).

Here the symbol \(r\rightarrow v\) means that \(f(r)=(v,v')\) for
some vertex \(v'\), or \(f(r)=(v,+)\).

 In
\(\bigotimes \limits_{(r\rightarrow v)}\tilde{\mathcal{H}}'\) the
tensor product is over all lines \(r\) such that \(r\rightarrow
v\). Let \(v\) be a vertex of the tree \(T\). If \(f(r)=(v',v)\)
for some vertex \(v'\) or \(f(r)=(v,+)\) we say that the line
comes from the vertex \(v\). If \(f(r)=(v,v')\) for some vertex
\(v'\) or \(f(r)=(v,-)\) we say that the line comes into the
vertex \(v\).

\textbf{Definition.} Let \((T,\varphi,\vec{\tau})\) be a tree of
correlations such that for each vertex \(v\)
\(\varphi(v)=\mathcal{L}_{int}^{c,l_v}\), where \(l_v\) is a
number of lines coming into \(v\). We call this tree the von
Neumann tree and denote it by \(T_{\vec{\tau}}\). We also say that
\(\varphi\) is a von Neumann vertex function.

\textbf{Definition.} To each tree of correlation
\((T,\varphi,\vec{\tau})\) assign an element
\begin{eqnarray}
U^t_{T,\varphi}(\vec{\tau}) {\in \rm Hom \mit}(\bigoplus
\limits_{R_{sh}} \tilde{\mathcal{H}}',\bigoplus \limits_{R_{root}}
\tilde{\mathcal{H}}')
\end{eqnarray}
by the following way:

If \(T\) is disconnected then
\begin{eqnarray}
U^t_{T,\varphi}(\vec{\tau}) f_1\otimes...\otimes f_n \nonumber\\
=\bigotimes \limits_{CT}\{
U^t_{CT,C\varphi}(C\vec{\tau})\bigotimes \limits_{i \in R_{sh}(CT)}
f_i\}.
\end{eqnarray}
Here the number of connected components of \(T\) is equal to \(n\)
and connected components of \(T\) are denoted by \(CT\).
\(C\varphi\) and \(C\vec{\tau}\) are the restrictions of
\(\varphi\) and \(\vec{\tau}\) to the sets of vertices and lines
of \(CT\) respectively. \(R_{sh}(CT)\) is a set of shoots of \(CT\).
Now let \(T\) be a connected tree. To define
\begin{eqnarray}
U^t_{T,\varphi}(\vec{\tau})\bigotimes \limits_{r \in R_{sh}} f_r
\end{eqnarray}
by induction its enough to consider the following two cases.

 case 1). The tree \(T\) has no shoots.

 a) Suppose that the tree \(T\) has more than one vertex. Let
 \(v_{min}\) be some minimal vertex of \(T\) and \(v_0\) be a
 vertex such that an unique line \(r_0\) comes from \(v_{min}\)
 into \(v_0\). Let \(T'\) be a tree obtained from \(T\) by
 removing the vertex \(v_{min}\) of \(T\). Let
 \(\vec{\tau}'\) be a restriction of \(\vec{\tau}\) to
 \(R\setminus\{r_0\}\). Let \(\varphi'\) be a function, defined on
 \(V\setminus\{v_{min}\}\) as follows: \(\varphi'(v)=\varphi(v)\)
 if \(v\neq v_0\) and
 \begin{eqnarray}
 \varphi'(v_0)\bigotimes \limits_{r\rightarrow v_0;\;
 r\neq r_0} f_r=\varphi(v_0)\bigotimes \limits_{r\rightarrow v_0} h_r,
 \end{eqnarray}
 where
 \begin{eqnarray}
 h_r=f_r\; \rm if \mit \; r\neq r_0,\; \rm and \mit \nonumber\\
 h_{r_0}=e^{\mathcal{L}_0\tau(r_0)} \varphi(v_{min}).
 \end{eqnarray}

 Put by definition
 \begin{eqnarray}
 U^t_{T,\varphi}(\vec{\tau})\rangle=U^t_{T',\varphi'}(\vec{\tau}')\rangle,\;
\end{eqnarray}

b) The tree \(T\) has only one vertex \(v_{min}\). Then
\begin{eqnarray}
U^t_{T,\varphi}(\vec{\tau})=e^{-(t-\tau)\mathcal{L}_0}\varphi(v_{min}).
\end{eqnarray}

Case 2.) The tree \(T\) has a shoot \(r_0\) coming into the vertex
\(v_0\). In this case instead the tree \((T,\varphi,\vec{\tau})\)
consider the tree \((T',\varphi',\vec{\tau}')\), where the tree
\(T'\) has the same vertices as \(T\), the set of lines of \(T\)
is obtained by removing the line \(r_0\) from the set of lines of
\(T'\), the function \(\vec{\tau}'\) is a restriction of the
function \(\vec{\tau}\) to the set of lines of \(T\) and the
function \(\varphi'\) is defined as follows:
\begin{eqnarray}
\varphi'(v)=\varphi(v), \rm if \mit v\neq v_0 \; \rm and \mit \nonumber\\
\varphi'(v_0)\bigotimes \limits_{r\rightarrow v;\;r\neq r_0} h_r=
\varphi(v_0)\bigotimes \limits_{r\rightarrow v;}
g_r,\;\rm where \mit \nonumber\\
g_r=h_r,\; \rm if \mit \;r\neq r_0,\; \rm and \mit \nonumber\\
g_r=e^{\mathcal{L}_0(t-t_r)} f_{r_0}.
\end{eqnarray}
Here we put
\begin{eqnarray}
t_r=\sum \tau_{r'},
\end{eqnarray}
where the sum is over all lines \(r'\) which forms decreasing way
coming from \(+\) to \(v_0\).
Put by definition
\begin{eqnarray}
U^t_{T,\varphi}(\vec{\tau})|f\rangle:=U^t_{T',\varphi'}(\vec{\tau'})|f'\rangle,
\end{eqnarray}
where
\begin{eqnarray}
|f'\rangle=\bigotimes \limits_{r \in (R_{sh})_{T'}} f_r.
\end{eqnarray}

Let \((T,\varphi,\vec{\tau})\) be some tree of correlations. We
can identify the tensor product
\begin{eqnarray}
\bigotimes \limits_{r \in R_{Sh}}\tilde{\mathcal{H}}'_r
\end{eqnarray}
with
\begin{eqnarray}
\bigotimes \limits_{i=1}^{sh(T)}\tilde{\mathcal{H}}'
\end{eqnarray}
and the tensor product
\begin{eqnarray}
\bigotimes \limits_{r \in R_{root}}\tilde{\mathcal{H}}'_r
\end{eqnarray}
with
\begin{eqnarray}
\bigotimes \limits_{i=1}^{r(T)}\tilde{\mathcal{H}}.'
\end{eqnarray}
Using these identifications consider an operator
\(V^t_{T,\varphi}(\vec{\tau}):\mathcal{H}^c\rightarrow
\mathcal{H}^c\) defined by the following formula
\begin{eqnarray}
V^t_{T,\varphi}=\rm sym \mit \circ U^t_{T,\varphi} \circ
P_{sh(T)},
\end{eqnarray}
where \(P_{sh(T)}\) is a projection of \(\mathcal{H}_c\) to \(\rm
sym \mit \bigotimes \limits_{i=1}^{sh(T)} \tilde{\mathcal{H}}'\).

\textbf{Remark.} If \((T,\varphi,\vec{\tau})\) is a von Neumann
tree of correlations then we will shortly denote the operators
\(U^t_{(T,\varphi)}\) and \(V^t_{(T,\varphi)}\) by \(U^t_T\) and
\(V^t_T\) respectively.

The following theorem holds:

\textbf{Theorem.} The following representation for the evolution
operators holds holds (in the sense of formal power series on
coupling constant \(\lambda\)).
\begin{eqnarray}
U^c(t',t'')=\sum \limits_T\frac{1}{n_T!} \int \limits_{\forall r
\in R_{sh}\;t-t_r>t''} V^t_T(\vec{\tau})d \vec{\tau}.
\end{eqnarray}
Here \(n_T\) is a number of vertices of the directed tree \(T\).
\section{The general theory of renormalization of
\(U(t,-\infty)\rangle\)}

In the present section we by using the decomposition of
correlation's dynamics by trees describe the general structure of
counterterms of \(U(t,-\infty)\rangle\), which subtract the
divergences from \(U(t,-\infty)\rangle\).

Let \(T\) be a tree. Let us give a definition of its right
subtree.

\textbf{Definition.} Let \(v_1,...,v_n\) be vertices of \(T\) such
that \(\forall i,j=1,...,n,\;i\neq j\) \(v_i\nless v_j\). Let us
define subtree \(T_{v_1,...,v_n}\). The set of vertices
\(V_{T_{v_1,...,v_n}}\) of \(T_{v_1,...,v_n}\) by definition
consists of all vertices \(v\) such that \(v\leq v_i\) for some
\(i=1,...,n\).

The set \(R_{T_{v_1,...,v_n}}\) of all lines of the tree
\(T_{v_1,...,v_n}\) consists of all lines \(r\) of \(R_T\) such
that such that \(h(r)=(v'',v')\) and \(v',v''\leqslant v_i\) for
some \(i=1,...,n\). The incident relations on \(T_{v_1,...,v_n}\)
are induced by the incident relations of \(T\) except the
following point: if the line \(r\) comes from the vertex \(v\)
into \(v_i,\;i=1,...,n\) we put
\(f_{T_{\{v_1,...,v_n\}}}(r)=(v,+)\). In this case the line \(r\)
is a root and the vertex \(v\) is a root vertex of the tree
\(V_{T_{\{v_1,...,v_n\}}}\). The tree \(T_{v_1,...,v_n}\) is
called a right subtree of \(T\).

\textbf{The Bogoliubov --- Parasiuk  renormalization
prescription.} Let us define the following operator:
\begin{eqnarray}
W_{r_0}(t)=\bigotimes \limits_{r \in R_{root}(T)}Z_{r,r_0}(t),
\end{eqnarray}
where by definition,
\begin{eqnarray}
Z_{r,r_0}(t)=1,\;\rm if \mit\; r\neq r_0,\;\rm and \mit \nonumber\\
 Z_{r,r_0}(t)=e^{-\mathcal{L}_0t}.
 \end{eqnarray}
 We say that the amplitude \(A_{T,\varphi}\) is time ---
 translation invariant amplitude if for each tree \(T\) and for
 each its root line \(r_0\)
 \begin{eqnarray}
 W_{r_0}(t)A_{T,\varphi}=A_{T,\varphi}.
 \end{eqnarray}

 For each set of amplitudes \(A_{T,\varphi}\) put by definition:
\begin{eqnarray}
A_{T,\varphi}\rangle=F\circ A_{T,\varphi},
\end{eqnarray}
where \(T\) is an arbitrary tree without shoots.

 The renormalization assign to each tree \(T\) without shoots the amplitudes
 \(\Lambda_{T,\varphi}\) satisfying to the following relations:

 a) If the tree \(T\) is not connected and \(\{CT\}\) is a set of
 its
 connected components, while \(\{C\varphi\}\) is a set of its
 a restriction of \(\varphi\) to \(CT\)
 \begin{eqnarray}
 \Lambda_{T,\varphi}=\bigotimes \limits \Lambda_{CT,C\varphi}
 \end{eqnarray}
 in obvious notations.

  b) The amplitudes \(\Lambda_{T,\varphi}\) are real i.e.
 \begin{eqnarray}
 (\Lambda_{T,\varphi})^\star=\Lambda_{T,\varphi^*}
 \end{eqnarray}

 c) The amplitude \(\Lambda_{T,\varphi}\)  satisfy to the property of
 time-translation invariance.

 It has been proven that
 \begin{eqnarray}
 U(t,-\infty)\rangle=\sum \limits_T \frac{1}{n_T} \int d\vec{\tau}
 U_T^t(\vec{\tau})\rangle. \label{63}
 \end{eqnarray}

In the last formula the summation is over all trees \(T\) without shoots.

 Let \(T\) be a tree without shoots and \(T'\) be a right subtree
 of \(T\) in the sense described before. Let us define the
 amplitude
 \begin{eqnarray}
 \Lambda_{T,\varphi}\star U^t_{T,\varphi}(\vec{\tau}).
 \end{eqnarray}
Let \(T\backslash T'\) by definition be a tree obtained by
removing from the set \(V_T\) all the vertices of \(T'\) and from
the set \(R_T\) all the internal lines of \(T'\). In (\ref{63})
\(\vec{\tau}\) is a map from \(R_{T\setminus T'}\) into
\(\mathbb{R}^+\).

We can consider the amplitude \(U^t_{T\setminus T'}\) as a map
\begin{eqnarray}
{\bigotimes \limits_{(R_{T\setminus
T'})_{sh}}}\tilde{H}'\rightarrow {\bigotimes
\limits_{(R_{T\setminus T'})_{root}}}\tilde{H}'.
\end{eqnarray}
By using this identification we simply put
\begin{eqnarray}
\Lambda_{T',\varphi}\star U^t_{T,\varphi}(\vec{\tau})\nonumber\\
=U^t_{T\setminus T',\varphi}(\vec{\tau})\Lambda_{T',\varphi}.
\end{eqnarray}
Now let us define the renormalized amplitudes, by means the
counterterms \(\Lambda_T\) by the following formula:
\begin{eqnarray}
(R_\Lambda U)(t,-\infty) \nonumber\\
=\sum \limits_T \frac{1}{n_T!}\sum \limits_{T'\subset T} \int
\Lambda_{T'}\star U_T^t(\vec{\tau}) d\vec{\tau}.
\end{eqnarray}

The renormalized amplitudes satisfy the following properties:

\textbf{Property 1.} For each \(t \in \mathbb{R}\)
\begin{eqnarray}
(R_\Lambda U)(t,-\infty)\rangle\nonumber\\
=e^{-\mathcal{L}_0 t}(R_\Lambda U)(0,-\infty)\rangle.
\end{eqnarray}
This property simply follows from the definition of \((R_\Lambda
U)(t,-\infty)\rangle\) and means that the state \((R_\Lambda
U)(t,-\infty)\rangle\) is a stationary state.

\textbf{Property 2.}

\begin{eqnarray}
(R_\Lambda U)(t,-\infty)\rangle=U(t,0)(R_\Lambda
U)(0,-\infty)\rangle.
\end{eqnarray}
This property follows from the following representation of \(
(R_\Lambda U)(t,-\infty)\rangle\).
\begin{eqnarray}
(R_\Lambda U)(t,-\infty)\rangle= U(t,-\infty)\mathcal{I}\rangle,
\end{eqnarray}
where
\begin{eqnarray}
\mathcal{I}\rangle= \sum \limits_{T}
\frac{1}{n_T!}\Lambda_T\rangle,
\end{eqnarray}
and the sum in the last formula is over all von Neumann trees
without shoots. Property 2 means that the state \((R_\Lambda
U)(t,-\infty)\rangle\) satisfy to the von Neumann dynamics.

Below we will prove that we can find the counterterms
\(\Lambda_T\) such that \((R_\Lambda U)(t,-\infty)\rangle\) is
finite.

We will prove also that the counterterms \(\{\Lambda_T\}\) satisfy
to the following property.

d) Let \(T\) be an arbitrary connected tree without shoots.
Consider the following element of \(\mathcal{H}'\):
\begin{eqnarray}
a:=\sum \limits_T \sum \limits_{T'\subset T} \int
\Lambda_{T'}\star U^t_T(\vec{\tau})\rangle d \vec{\tau}.
\end{eqnarray}
We can represent the element \(a\) as follows:
\begin{eqnarray}
a=\sum \limits_{k,l,f,g=0}^{\infty} \int
w_m(x_1,...,x_{k}|y_1,...,y_{l}|v_1,...,v_{f}|w_1,...,w_{g})\nonumber\\
:\prod \limits_{i=1}^{k_m} a_+^+(x_i)dx_i \prod
\limits_{i=1}^{l_m} a_+(y_i)d y_i \prod \limits_{i=1}^{f_m}
a_-(v_i)d v_i\prod \limits_{i=1}^{g_m} a_-^+(w_i) d w_i:.
\end{eqnarray}
Let \(\tilde{w}_{k,l,f,g}(z_1,...,z_n)\) (\(n=k_m+l_m+f_m+g_m\))
be a Fourier transform of
\(w_{k,l,f,g}(x_1,...,x_{k}|y_1,...,y_{l}|v_1,...,v_{f}|w_1,...,w_{g})\).
Then
\begin{eqnarray}
 \int dz_1,...,dz_n
\tilde{w}_{k,l,f,g}(z_1+s(1)e_1a,...,z_n+s(n)e_1a)f(z_1,...,z_n),
\end{eqnarray}
tends to zero faster than ever inverse polynomial on \(a\) as
\(a\rightarrow +\infty\). Here \(s(i)\) are the numbers from
\(\{0,1\}\) and there exist numbers \(i,j,\;i,j=1,...,n\) such
that \(s(i)=0,\;s(j)=1\) for some \(i,j=1,...,n\).
\(f(z_1,...,z_n)\) is a test function. \(e_1\) is a unite vector parallel to the \(x\)-axis.

\textbf{Remark.} The property d) implies the weak cluster property
of the state
\begin{eqnarray}
(RU)(0,-\infty)\rangle.
\end{eqnarray}

\section{The Friedrichs diagrams}

Now let us start to give a constructive description of the
counterterms \(\Lambda_T\) such that the
amplitude \(R(U)(t,-\infty)\rangle\) is finite, and the
counterterms \(\Lambda_T\) satisfy the properties a) --- d) from
the previous section.

At first we represent \(U^t_{T,\varphi}(\vec{\tau})\), where \(T\)
is some tree without shoots, as a sum over all so-called
Friedrichs graphs \(\Phi\) concerned with \(T\).

\textbf{Definition.} A Friedrichs graph \(\Phi_T\) concerted with
 the directed tree \(T\) without shoots is a set
 \((\tilde{V},R,Or,f^+,f^-,g)\), where \(\tilde{V}\) is a union of the set of vertices of
 \(T\) and the set \(\{\oplus\}\). Recall that there is a partial
 order on \(V_T\). We define a partial order on the set \(\tilde{V}\) if
 we put \(\forall v \in V_T\) \(\oplus> v\). \(f^+\) and \(f^-\)
 are the maps \(f^+,f^-:R\rightarrow V\) such that \(f^+(r)>f^{-}(r)\).
 \(Or\) is a map \(R\rightarrow \{+,-\}\) called an orientation.
 \(g\) is a function which to each pair \((v,r)\), \(v \in V_T\),
 \(r \in R\) such that \(f^+(r)=v\) or \(f^-(r)=v\) assigns \(+\)
 or \(-\). The graph \((V,R,Or,f^+,f^-,g)\) must satisfy to the
 property: if we consider \(\oplus\) as a vertex, the obtained
 graph is connected.

 If \(f^+(r)=v\) we will write \(r\rightarrow v\), and if \(f^-(r)=v\) we will write \(r\leftarrow
 v\).

If we want to point out that the object \(B\) concerned with the
graph \(\Phi\) we will write \(B_\Phi\). For example we will write
\(V_\Phi\) and \(R_\Phi\) for the sets of vertices and lines of
\(\Phi\) respectively.

 At the picture we will represent the elements of \(V\) by points and
 the element \(\oplus\) by \(\oplus\). We will represent the elements
 of \(R\) by lines. The line \(r\) connects the vertices
 \(f^+(r)\) and \(f^{-}(r)\) at the picture. We will represent
 orientation \(Or(r)\) by arrow on \(r\). If \(Or(r)=+\) the
 arrow oriented from \(f^-(r)\) to \(f^+(r)\). If \(Or(r)=-\) the
 arrow oriented from \(f^+(r)\) to \(f^-(r)\). To represent the
 map \(g:(r,v)\rightarrow \{+,-\}\) we will draw the symbol
\(g((r,v))\) (\(+\)
 or \(-\)) near each shoot \((r,v)\). At the picture a shoot \((r,v)\) is a
 small segment of the line \(r\) near \(v\).

 \textbf{Definition.} The Friedrichs diagram \(\Gamma\) is a set
 \((T,\Phi,\varphi,h)\), where \(T\) is a tree, \(\Phi\) is a
 Friedrichs graph, \(\varphi\) is a map which assign to each
 vertex \(v\) of \(T\) a function of momenta \(\{p_r|
 r \in R_\Phi \}\) of the form
 \begin{eqnarray}
\varphi_v(...p_{r\leftrightarrows
v}...)=\psi_v(...p_{r\leftrightarrows v}...)\prod
\limits_{S_i}\delta(\sum \limits_{j=1}^{j_i} \pm p_i^j),
 \end{eqnarray}
where \(\psi_v\) is a test function of momenta coming into (from)
the vertex \(v\). \(\{S_i\}_{i=1}^{n_v}\) is a decomposition of
the set of shoots of \(v\) into \(n_v\) disjunctive nonempty sets
\(S_i\), \(p_i^1,...,p_i^{j_i}\) are momenta corresponding to the
shoots from \(\{S_i\}\), \(h\) is a function which assign to each
pair \(v \in V\), \(r \in R\) such that \(f^+(r)>v>f_-(r)\) a real
positive number \(h(v,r)\).

It will be clear that it is enough to consider only the diagrams \(\Gamma\) such that
for each its vertex \(v\) and set \(S_i\in \{S_i\}_{i=1}^{n_v}\) there exists a line \(r\) such that
\((r,f^{-}(r))\in S_i\).

To each Friedrichs diagram \(\Gamma=(T,\Phi,\varphi)\) assign
 an element of \(\tilde{\mathcal{H}}''_c\) of the form
 \begin{eqnarray}
 U^t_{(T,\Phi,\varphi)}(\vec{\tau})\nonumber\\
 =\int ...dp_{ext}...U^t_\Gamma(...p_{r_{ext}}...)\\
 \times :...a^{\pm}_{\pm}(p_{r_{ext}})...:\rangle.
 \end{eqnarray}
 Here \(p_{r_{ext}}\) are momenta of external lines, i.e. such lines
 \(r\) that \(f^+(r)=\oplus\). We chose the lower index of
 \(a^{\pm}_{\pm}(p_{r_{ext}})\)
 by the following rule.
 Let \(v\) be a vertex such that \(f^-(r_{ext})=v\). If
 \(g((r,v))=+\) we chose \(+\) as a lower index, and if
\(g((r,v))=-\) we chose \(-\) as a lower index. We chose the upper
index of \(a^{\pm}_{\pm}(p_{r_{ext}})\) by the following rule. If
the lower index of \(a^{\pm}_{\pm}(p_{r_{ext}})\) is \{-\} then
the upper index is equal \(+\) if the corresponding line comes
from the vertex \(v\) and this index is equal \(-\) if the
corresponding line comes into the vertex \(v\). If the lower index
of \(a^{\pm}_{\pm}(p_{r_{ext}})\) is \{+\} then the upper index is
equal \(-\) if the corresponding line comes from the vertex \(v\)
and this index is equal \(+\) if the corresponding line comes into
the vertex \(v\).

Now let us describe the amplitude \(U_\Gamma^t(...p_{ext}...)\).
By definition we have
\begin{eqnarray}
U^0_\Gamma(\vec{\tau})(...p_{ext}...)\nonumber\\
=\int \limits_{r \in R_{in}} \prod \limits_{v} \varphi_v(...
p_{r\leftrightarrows v}...)  \nonumber\\
\times \prod \limits_{r \in R_\Gamma} e^{i Or(r) p_r^2 (\sum
\limits_{r_T \in (R_T)_r} \tau_{r_T}+\sum \limits_{v \in
V_r}h(v,r))}dp_r \nonumber\\
\prod \limits_{r \in R} G(Or(r), g(f^+(r)),g(f^-(r)))(p).
\label{77}
\end{eqnarray}

Let us describe the elements of this formula. \(R_\Gamma\) is a set of
all lines of diagram \(\Gamma\). Symbol \(r\leftrightarrows v\)
denotes that the line \(r\) comes into (from) the vertex \(v\). In
the expression
\begin{eqnarray}
\psi_v(... p_{r\leftrightarrows v}...)\delta(\sum
\limits_{r\leftrightarrows v} \pm p_r)
\end{eqnarray}
we take the upper sign \(+\) if the line \(r\) comes into the
vertex \(v\) and we take lower sign \(-\) in the opposite case.
The symbol \(R_T\) denotes the set of lines of the tree \(T\) from
the triple \((T,\Phi,\varphi)\) and symbol \(r_T\) means the line
from \(R_T\). The symbol \(V_r\) denotes the set of all vertices
\(v\) such that \(f^+(r)\geq v\geq f^-(r)\). The symbol \((R_T)_r\)
denotes the set of all lines \(r_T\) of \(R_T\) such that the
increasing path coming from \(f^-(r)\) into \(f^+(r)\) contains
\(r_T\). \(G(Or(r),g(f^+(r)),g(f^-(r)))(p)\) is a factor defined
as follows
\begin{eqnarray}
G(Or(r),g(f^+(r)),g(f^-(r)))(p)\delta(p-p')\nonumber\\
=\rho_0'(a^{sgn(-Or(r)g((r,f^+(r)))}_{g((r,f^+(r)))}(p),a^{sgn(Or(r)g((r,f^-(r)))}_{g((r,f^-(r)))}(p')).
\end{eqnarray}
Below we will simply write \(G_r(p)\) instead of
\(G(Or(r),g(f^+(r)),g(f^-(r)))(p)\).

It is evident that we can represent \(U^0_T(\vec{\tau})\) as a sum
over some Friedrichs diagrams \(\Gamma\) corresponding to the tree
\(T\) of the quantities \(U^0_\Gamma (\vec{s})\).

Now let us define the quotient diagrams.

\textbf{Definition}. Let \(\Gamma=(T,\Phi,\varphi,h)\) be a
Friedrichs diagram and \(A\subset R_T\) be a subset of the set
\(R_T\) of lines of \(T\) and \(\vec{\tau}\) be a map from \(R_T\)
into \(\mathbb{R}^+\).

We define the quotient diagram
\(\Gamma_{A\vec{\tau}}:=(T_A,\Phi_A,\varphi_{A\vec{\tau}},h_A)\)
by the following way. To obtain the tree \(T_A\) we must tighten
all lines from \(A\) into the point. To obtain \(\Phi_A\) we must
remove all loops obtained by tightening all lines from \(A\) into
the point.

Now let us define \(\varphi_{A\vec{\tau}}\). Joint all the
vertices of \(T\) to \(A\). We obtain a tree denoted by \({}^AT\).
Let \(\{C{}^AT\}\) be a set of all connected components of
\({}^AT\). Let \(v_0\) be a vertex of \(\Phi_A\) corresponding to
the connected component \(C{}^AT\) of \({}^AT\). Put by
definition:
\begin{eqnarray}
\varphi_\Gamma(...p_{r\leftrightarrows
v}...)_{A\vec{\tau}}\nonumber\\
=\int \prod_{r \in R_{in}} \prod \limits_{v \in V}\varphi_v(...\pm
p_{r\leftrightarrows v}...) \nonumber\\
\times \prod \limits_{r \in R_{in}} e^{iOr(r)p_r^2(\sum \limits_{r
\in (R_T)_r} \tau_{r_T}+\sum \limits_{v \in V_r} h(v,r))}.
\end{eqnarray}

Let us point out that is that in the previous formula. \(R_{in}\)
is a set off all lines of \(\Phi_A\) such that \(f^+(r)\) and
\(f^-(r)\) are the vertices of \(C{}^AT\). \((R_T)_r\) denotes the
set of all lines \(r_T\) of \(R_T\) such that the increasing path
coming from \(f^-(r)\) into \(f^+(r)\) contains \(r_T\). The
symbol \(V_r\) denotes the set of all vertices \(v\) such that
\(f^+(r)\geq v\geq f^-(r)\). \(h_A(v_0,r)=\sum \limits_{v \in
V_{C{}^AT}}h(v,r)+\sum \limits_{r_T \in A;\; r_T \in (R_T)_r}
\tau_{r_T}\).

\textbf{Definition.} Let \(\Gamma\) be a Friedrichs diagram. Let
\(\mathcal{F}_\Gamma\) be a space of all functions of external
momenta of the diagram \(\Gamma\) of the form:
\begin{eqnarray}
 \psi(...p_{ext}...),
\end{eqnarray}
where \(\psi(...p_{ext}...)\) is a test function of external momenta.

The convolution of the amplitude
\(A_\Gamma(\vec{\tau})(...p_{ext}...)\) with the function \(f \in
\mathcal{F}_\Gamma\) we denote by \(A_\Gamma(\vec{\tau})[f]\).

\section{The Bogoliubov --- Parasiuk renormalization
prescriptions}

Let for each Friedrichs diagram \(\Gamma=(T,\Phi,\varphi)\)
\(A_\Gamma(\vec{\tau})(...p_{ext}...)\) be some amplitude. Fix
some diagram \(\Gamma\) and let \(T'\) be some right subtree of
the tree \(T\) corresponding to \(\Gamma\). Let \(\Gamma_{T'}\) be
a restriction of the diagram \(\Gamma\) on \(T'\) in obvious
sense. Define the amplitude \(A_{\Gamma_{T'}}\star U_\Gamma
(...p_{ext}...)\) by the following formula:
\begin{eqnarray}
A_{\Gamma_{T'}}\star U_\Gamma (...p_{ext}...)\nonumber\\
=\int \prod \limits_{r \in R'}\{e^{iOr(r)p_r^2(\sum \limits_{r_T \in
(R'_T)_r}\tau_{r_T}+\sum \limits_{v \in V'_r}h(v,r))}\}\prod
\limits_{v \in
V'}\varphi_v(...p_{r\rightleftarrows v}...)\nonumber\\
\times A_{\Gamma_{T'}}(...p...).
\end{eqnarray}
In this formula \(V'\) is a set of all vertices \(v\) such that
\(v\) is not a vertex of \(V_{T'}\), \(R'\) is a set of all lines
\(r\) of \(\Phi_\Gamma\) such that \(f^+(r)\) is not a vertex of
\(T'\).
\((R_T')_r\) is a set of all lines \(r_T\) of \(T\) such that
\(r_T\) is not a line of \(T'\) and there exists an increasing
path on \(T\) coming from \(f^{-}(r)\) into \(f^+(r)\) such that
this path contains \(r_T\). \(V'_r\) is a set of all vertices
\(v\) of T such that v is not a vertex of \(T'\) and
\(f^+(r)\geq v\geq f^-(r)\).

Let \(A_\Gamma(\vec{\tau})(p)\) be some amplitude. Put by
definition:
\begin{eqnarray}
\hat{A}_\Gamma(s_1,...,s_n)(p):=A_\Gamma(\frac{1}{s_1},...,\frac{1}{s_n})(p)\prod
\limits_{i=1}^{n} \frac{1}{s_i^2},
\end{eqnarray}
where \(n\) is a number of lines of \(T_\Gamma\). Below we will
consider the amplitudes \(\hat{A}_\Gamma(\vec{s})[f]\) as
distributions on \((\mathbb{R}^+)^n\) i.e. as an element of the
space of tempered distributions \(S'((\mathbb{R}^+)^n)\). Let
\(\psi(\vec{s})\) be a test function from \(S((\mathbb{R}^+)^n)\).
The convolution of the amplitude \(\hat{A}_\Gamma(\vec{s})[f]\)
and the function \(\psi(\vec{s})\) we denote by:
\begin{eqnarray}
\langle\hat{A}_\Gamma(\vec{s})[f],\psi(\vec{s})\rangle\nonumber\\
:=\int \limits_{(\mathbb{R}^+)^n} d\vec{s}
\hat{A}_\Gamma(\vec{s})[f]\psi(\vec{s}).
\end{eqnarray}

\textbf{The Bogoliubov --- Parasiuk prescriptions.} It will be clear below,
that we can take into account only the diagrams \(\Gamma\) such that for each line \(r_T\)
of the corresponding tree of correlations \(T\) \(\sharp R_{r_T}\geq 3\). Here \(R_T\) is a set of all lines \(r\) of
\(\Gamma\) such that the increasing path on \(T\) which connects \(f^-(r)\) and \(f^+(r)\) contain \(r_T\).
Other diagram can be simply subtracted by the counterterms
\(\Lambda_T\). Below we will consider only such diagrams.

According to the Bogoliubov --- Parasiuk prescriptions we must to
each diagram \(\Gamma\) assign the counterterm amplitude
\(\hat{C}_\Gamma(\vec{s})[f]\) \(f \in \mathcal{F}_\Gamma\)
satisfying the following properties.

a) (Locality.) \(\hat{C}_\Gamma(\vec{s})[f]\) is a finite linear
combination of \(\delta\) functions centered at zero and their
derivatives.

b) Let \(\Gamma\) be a Friedrichs diagram and \(T\) be a
corresponding tree of correlations. Let \(A \subseteq R_T\) and
\(T'\) is some right subtree of \(T\) such that:

1) all lines \(r_T\) of T such that \(r_T\) is not a line of
\(R_{T'}\) belongs to \(A\),

2) All the root lines of \(T'\) do not belongs to \(A\).

Then
\begin{eqnarray}
\hat{C}_{\Gamma_{A\vec{\tau}}}(\vec{s})[f]=(\hat{C}_{\Gamma_{A'\vec{\tau}}'}\star
\hat{U}_\Gamma)(\vec{s})[f],
\end{eqnarray}
where \(A':=A\cap (R_{T'})\) and \(\Gamma'\) is a restriction of
\(\Gamma\) on \(T'\).

c)
\begin{eqnarray}
\hat{C}_\Gamma(\vec{s})[f]=-\mathbb{T}(\sum
\limits_{\emptyset\subset A\subset R_{T_\Gamma}}
\hat{C}_{\Gamma_{A\vec{\tau}}}(\vec{s})[f]+\hat{U}_{\Gamma}(\vec{s})[f]),
\end{eqnarray}
where
\(\vec{\tau}=(\tau_1,...,\tau_n)=(\frac{1}{s_1},...,\frac{1}{s_n})\),
the symbol \(\subset\) means the strong inclusion and
\(\mathbb{T}\) is some subtract operator.

d) The amplitudes \(\hat{C}_\Gamma(\vec{s})[f]\) satisfy to the property
of time-translation invariance, i.e.
\begin{eqnarray}
e^{i\sum \limits_{r \in (R_{root})_\Gamma} Or(r) p_r^2 t}
\hat{C}_\Gamma(\tau_1,...,\tau_n)[f] =\hat{C}_\Gamma(\tau_1+t,...,\tau_n)[f].
\end{eqnarray}

e) Let \(\Gamma\) be a Friedrichs diagram. Let
\begin{eqnarray}
\hat{R}_\Gamma'(\vec{s})[f]:=\hat{U}_\Gamma(\vec{s})[f]+\sum
\limits_{\emptyset \subset A\subset R_{T_\Gamma}}
\hat{C}_{\Gamma_{A\vec{\tau}}}(\vec{s})[f],\; \rm and \mit \nonumber\\
\hat{R}_\Gamma(\vec{s})[f]:=\hat{U}_\Gamma(\vec{s})[f]+\sum
\limits_{\emptyset \subset A\subset R_{T_\Gamma}}
\hat{C}_{\Gamma_{A\vec{\tau}}}(\vec{s})[f]+\hat{C}_\Gamma(\vec{s})[f].
\end{eqnarray}
The amplitudes \(\hat{R}_\Gamma(\vec{s})\) is well defined
distribution on \((\mathbb{R}^+)^n\).

f) The amplitudes \(\hat{R}_\Gamma(\vec{s})\) satisfy to the weak
cluster property. This property means the following. Let
\(f(...p_{ext}...)\) be a test function. Then
\begin{eqnarray}
\int dp \hat{R}_\Gamma(\vec{s})(...p_{ext}...) f(...p_{ext}...)
e^{ia\sum \limits_{r \in A} p_r^1}\rightarrow 0,
\end{eqnarray}
as \(a\rightarrow \infty\). Here \(p_r^1\) is a projection of \(p_r\) to the \(x\)-axis.
Note that the weak cluster property admit to prove the Gibbs formula if the system has not first
integrals except momenta and energy by the usual way.

For each diagram \(\Gamma\) put by definition:
\begin{eqnarray}
\Lambda_\Gamma(\vec{\tau})=\sum \limits_{A\subseteq R_{T_\Gamma}}'
C_{\Gamma_{A\vec{\tau}}},
\end{eqnarray}
where \('\) in the sum means that all the root lines of
\(T_\Gamma\) do not belong to \(A\).

Put
\begin{eqnarray}
\Lambda_T=\sum \limits_{\Gamma\sim T}
\int
\limits_{(\mathbb{R}^+)^n}d\vec{\tau}\Lambda_\Gamma(\vec{\tau})(...p_{ext}...)...a^\pm_\pm(p_{ext})...\rangle,
\end{eqnarray}
where the symbol \(\Gamma\sim T\) means that the sum is over all
diagrams corresponding to \(T\) with suitable combinatoric
factors. Suppose that the properties a) --- f) are satisfied. Then
\(\Lambda_T\) are the counterterms needed in the section .

\textbf{Theorem --- Construction.} It is possible to find such a
subtract operator \(\mathbb{T}\) such that there exists
counterterms \(\hat{C}_\Gamma\) satisfying to the properties a)
--- f).

Note that we can use not real counterterms. Indeed the evolution
operator is real, so after renormalization we can simply take
\(\rm Re \mit\; U(t,-\infty)\rangle\).

Before we prove our theorem let us prove the following

\textbf{Lemma.} Let \(L_1=S(\mathbb{R}^{k})\),
\(L_2=S((\mathbb{R}^+)^n)\), \(k,n=1,2,...\). Let \(A(p)\) be some
quadratic form on \(\mathbb{R}^{k}\). Let \(T^1_t\), \(t\geq 0\)
be a one-parameter semigroup acting in \(L_1\) as follows:
\begin{eqnarray}
T^1_t:f(...p...)\mapsto e^{iA(p)t}f(...p...).
\end{eqnarray}

Let \(T^2_t\) \(t\geq 0\) be some infinitely differentiable
semigroup of continuous operators in \(L_2\).

Let \(M\) be a subspace of finite codimension in \(L_2\). Suppose
that \(M\) is invariant under the action of \(T^2_t\), i.e.
\(\forall t>0\) \(T^2_t M\subset M\).

Suppose that there exist the linear independent vectors \(f_1,...f_l\) in \(L_2\)
such that

\begin{eqnarray}
\rm Lin \mit \{\{f_1,...,f_l\},
M\}=L_2,\nonumber\\
 M\cap \rm Lin \mit \{f_1,...,f_l\}=0.
 \end{eqnarray}
and for each \(i=1,...,l\), \(t\geq 0\)
\begin{eqnarray}
T_t^2 f_i=f_i+a_{i-1}f_{i-1}+...+a_1 f_1+f,
\end{eqnarray}
for some coefficients \(a_{i-1},...,a_1\) and the element \(f \in
M\).

Let \(g\) be a functional on \(L_1\otimes M\) such that \(g\) is
continuous with respect to the topology on \(S(\mathbb{R}^k)\times
S((\mathbb{R}^+)^n)\). Suppose that \(\forall f \in L_1\otimes M\)
and \(\forall t>0\) \(\langle g, T_t f\rangle=\langle g,f
\rangle\) where \(T_t=T^1_t\otimes T^2_t\).

Then there exists an continuous extension \(\tilde{g}\) of \(g\)
on \(S(\mathbb{R}^k)\times S((\mathbb{R}^+)^n)\) such that
\(\forall f \in L_1\otimes L_2\) and \(t>0\) \(\langle
\tilde{g},T_t f\rangle=\langle\tilde{g},f\rangle\).

By definition we say that the functional \(h\) on \(L_1\otimes
L_2\) is invariant if \(\forall t>0\) and \(\forall f \in
L_1\otimes L_2\) \(\langle h,T_t f\rangle=\langle h,f\rangle\).

\textbf{Proof of the lemma.} At first we extend our functional
\(g\) to the invariant functional \(\tilde{g}\) on \(L_1\otimes
L_2\) and then we prove that \(\tilde{g}\) is continuous.

Let \(N\) be a subspace of \(L_1\) of all functions of the form,
\(A(p)f(p)\), where \(f(p)\) is a test function. Let \(M_1=\rm Lin
\mit \{M\cup \{f_1\}\}\). Let

\begin{eqnarray}
h:=(\frac{d}{dt}T^2_t)|_{t=0} f_1.
\end{eqnarray}
Let \(k\) be a continuous functional on \(N\) defined as follows:
\begin{eqnarray}
\langle k,\varphi(p)A(p)\rangle=-\langle g,\varphi(p)\otimes
h\rangle. \label{ZUZU}
\end{eqnarray}
Let \(\tilde{k}\) be an arbitrary continuous extension of \(k\) on
whole \(L_1\). The existence of such continuation follows from Malgrange's preparation theorem \cite{5}.
 Now we define the continuous functional
\(\tilde{g}_1\) on \(L_1\otimes M_1\) as follows:
\begin{eqnarray}
\tilde{g}_1|_{L_1\otimes M}=g|_{L_1\otimes M},\nonumber\\
\langle \tilde{g}_1,f\otimes
f_1\rangle=\langle\tilde{k},f\rangle,\; \forall f \in L_1.
 \end{eqnarray}

 According to (\ref{ZUZU}) we find that \(\tilde{g}_1\) is an
 invariant extension of \(g\) on \(L_1\otimes M_1\). By the
 same method step by step we can extend the functional \(g\) to
 the functionals \(\tilde{g}_2\),
 ....\(\tilde{g}_l\) on \(L_1\otimes M_1\),...,\(L_l\otimes M_l\)
 respectively, where
 \( M_2= \rm Lin
\mit \{M\cup \{f_1,f_2\}\}\),...,\(M_l=\rm Lin \mit \{M\cup
\{f_1,f_2,...,f_l\}\}\) respectively. Just constructed functional
is separately continuous so it is continuous. The lemma is proved.

\textbf{Sketch of the proof of the theorem.} We will prove the theorem by
induction on the number of lines of the tree of correlation
\(T_\Gamma\) corresponding to the diagram \(\Gamma\).

The base of induction is evident. Suppose that the theorem is
proved for all diagrams of order \(<n\). (Order is a number of
lines of the tree of correlation.)

Let us give some definitions. Let \(\xi(t)\) be a smooth function
on \([0,+\infty)\) such that \(0\leq \xi(t)\leq 1\), \(\xi(t)=1\)
in some small neighborhood of zero and \(\xi(t)=0\) if
\(t>\frac{1}{3n}\). Let us define a decomposition of unite
\(\{\eta_A(\vec{s})|A\subset \{1,...,n\}\}\) by the formula
\begin{eqnarray}
\eta_A(\vec{s})=\prod \limits_{i \notin A} \xi(s_i) \prod
\limits_{i \in A}(1-\xi(s_i)).
\end{eqnarray}
We identify the set of lines of the tree of correlation with
\(\{1,...,n\}\) such that the root vertex is a vertex corresponds to 1.
Let \(\psi(x)\) be some test function on real
line such that \(\psi(t)\geq 0\), \(\int \psi(t)dt=1\) and
\(\psi(t)=0\) if \(|t|>\frac{1}{10}\). Put by definition:
\begin{eqnarray}
\delta_\lambda(x)=\frac{x}{\lambda^2}\psi(\frac{x}{\lambda}).
\end{eqnarray}
We have
\begin{eqnarray}
\int \limits_0^{+\infty}d \lambda \delta_\lambda(x-\lambda)=1.
\end{eqnarray}
Let \(S_N((\mathbb{R}^+)^n)\), \(N=1,2,...,\) be a subspace of
\(S((\mathbb{R}^+)^n)\) of all functions \(f\) such that \(f\) has
a zero at zero of order \(\geq N\).

 We have:
\begin{eqnarray}
\langle\hat{R}_{\Gamma}(\vec{s}\}[f],\Psi(\vec{s}) \rangle \nonumber\\
=\sum \limits_{A\subset \{1,...,n\}} \int \limits_0^{+\infty}d
\lambda \lambda^{n-1} \int \limits_{(\mathbb{R}^+)^n} d\vec{s}
\hat{R}_{\Gamma_{A \lambda
\vec{s}}}(\vec{s}|_{\{1,...n\}\setminus A} )\nonumber\\
\delta_1(1-|\vec{s}|)\Psi(\lambda\vec{s})\eta_A(\vec{s}) .
\label{GF}
\end{eqnarray}

The inner integral in (\ref{GF}) converges according to the
inductive assumption. Therefore if \(\Psi(\vec{s}) \in
S_N((\mathbb{R}^+)^n)\) and \(N\) is enough large the integral at
the right hand side of (\ref{GF}) converges. So
\begin{eqnarray}
\langle\hat{R}_{\Gamma}(\vec{s} )[f],\Psi(\vec{s}) \rangle
\end{eqnarray}
define a separately continuous functional on
\(S(\mathbb{R}^{3f})\otimes S_N((\mathbb{R}_+)^n)\). \(f=l-1\), where \(l\) is a number of
 external lines of \(\Gamma\). To define a subtract
operator \(\mathbb{T}\) we must to extend the functional
\(\langle\hat{R}_{\Gamma}(\vec{s})[f],\Psi(\vec{s}) \rangle\) to
the space \(S(\mathbb{R}^{3f})\otimes S((\mathbb{R})^n)\) such
that extended functional will satisfy to time-translation
invariant property. To obtain this extension we use the lemma. In
our case \(L_1=S(\mathbb{R}^{3f})\), \(L_2=S((\mathbb{R}^+)^n)\),
\(A(p)=\sum \limits_{r \in R_{ext}} Or(r)p_r^2\). \(T_t^2\) is an
operator adjoint to the following operator \(T_t^{2*}\) in the
\(S((\mathbb{R}_+)^n)\).
\begin{eqnarray}
T_t^{2*} f(s_1,...,s_n)=f(\frac{s_1}{1-s_1t},s_2,...,s_n)\; \rm if
\mit s_i<\frac{1}{t},\nonumber\\
T_t^{2*} f(s_1,...,s_n)=0,\; \rm if \mit s_i\geq \frac{1}{t}.
\end{eqnarray}
The basis \(\{f_1,...f_l\}\) from the lemma is
\(\{s_1^{m_1}....s^{m_n}_n\eta_{\{1,...,n\}}(\vec{s})\}\),
\(m_1,...m_n=1,2,3...\), \(m_1+m_2+...+m_n\leq N\)
lexicographically ordered. Now we can directly apply our lemma.

Now let us prove the weak cluster property. Let \(p \in\mathbf{R}^3\). Denote by
\(p^1,\;p^2,\;p^3\) the projections of \(p\) to the \(x,\;y,\;z\)-axis respectively. To prove the weak
cluster property it is enough to prove the following statement: for each connected diagram \(\Gamma\)
the function \(\langle F_\Gamma(\vec{s})(...p_{ext}...),\Psi(\vec{s})\rangle\) such that
\begin{eqnarray}
\delta(\sum \pm p_{ext})\langle F_\Gamma(\vec{s})(...p_{ext}...),\Psi(\vec{s})\rangle=
\langle\hat{R}_\Gamma(\vec{s})(...p_{ext}...),\Psi(\vec{s})\rangle
\end{eqnarray}
is a distribution on variables \(...p_{ext}^2...p_{ext}^3...\) (constrained by
momenta conservation low) which depends on \(...p_{ext}^1...\) (constrained by
momenta conservation low) by the differentiable way. We will prove this
statement by induction on the number of lines of the corresponding correlation's tree. The base of induction
is evident. Suppose that the statement is proved for all the correlation's trees such that the number of its
lines \(<n\). Let \(\Gamma\) be a diagram such that the number of the lines of the corresponding correlation's
 tree is equal to \(n\). It is evident that if \(\Psi(\vec{s})\) has a zero of enough large order at zero then
\(\langle\hat{F}_\Gamma(\vec{s})(...p_{ext}...),\Psi(\vec{s})\rangle\) belongs to the required class (its enough
to use our construction with decomposition of unite). Therefore we need to solve by induction the system of
equations of the form:
\begin{eqnarray}
(i\sum \pm (p_{ext})^2)\langle F_\Gamma(\vec{s})(...p_{ext}...),\Psi(\vec{s})\rangle\nonumber\\
=\langle F_\Gamma(\vec{s})(...p_{ext}...),\frac{d}{dt}T_t^{2*}\Psi(\vec{s})\rangle.
\end{eqnarray}
According to Malgrange's \cite{5} preparation theorem we can chose the solution \(\langle F_\Gamma(\vec{s})(...p_{ext}...)\) such that it belongs to the required class if
\( \langle F_\Gamma(\vec{s})(...p_{ext}...),\frac{d}{dt}T_t^{2*}\Psi(\vec{s})\rangle\) belongs to the
required class. Therefore the statement is proved. So our theorem is proved.

\section{Conclusion} In the present theory we have developed the
general theory of the renormalization of nonequlibrium diagram
technique. To study this problem we have used some ideas of the
theory of \(R\)-operation developed by N.N. Bogoliubov and O.S.
Parasiuk. This paper is formally independent of the previous paper
of this series. But the previous paper can be considered as an
illustration of some technical aspects of our theory that have
been omitted in the present paper which ends our series.

Author is grateful to Yu. E. Lozovik and I.L. Kurbakov for very useful discussions.

\end{document}